\documentclass[prd,twocolumn,nofootinbib,aps,superscriptaddress,tightenlines,preprintnumbers]{revtex4}
\usepackage{epsfig,latexsym,amssymb}

\def\tr{\text{Tr}\,}

\newcommand{\be}{\begin{equation}}
\newcommand{\ee}{\end{equation}}
\newcommand{\bea}{\begin{eqnarray}}
\newcommand{\eea}{\end{eqnarray}}
\newcommand{\mn}{{\mu\nu}}
\newcommand{\hb}{\bar{h}}
\newcommand{\call}{\mathcal{L}}
\newcommand{\mpl}{M_P}

\begin{document}

\preprint{\hbox{CALT-68-2670}  } 

\title{Aether Compactification} 

\author{Sean M. Carroll and Heywood Tam}

\affiliation{California Institute of Technology, Pasadena, CA 91125 }

\begin{abstract}
We propose a new way to hide large extra dimensions without invoking branes, based on 
Lorentz-violating tensor fields with expectation values along the extra directions.  We investigate 
the case of a single vector ``aether'' field on a compact circle.  In such a background, interactions
of other fields with the aether can lead to modified dispersion relations, increasing the
mass of the Kaluza-Klein excitations.  The mass scale characterizing each Kaluza-Klein tower
can be chosen independently for each species of scalar, fermion, or gauge boson.
No small-scale deviations from the inverse square law for gravity are predicted, although light
graviton modes may exist. 
\end{abstract}

%\date\today
\maketitle

%%%%%%%%%%%%%%%%%%%%%%%%%%%%%%%%%%%%%%%%%%%%%
\section{Introduction}

If spacetime has extra dimensions in addition to the four we perceive, they are somehow
hidden from us.   For a long time, the only known way to achieve this goal was the classic
Kaluza-Klein scenario:  compactify the dimensions on a manifold of characteristic size
$\sim R$.  Momentum in the extra dimensions is then quantized in units of $R^{-1}$, giving rise to 
a Kaluza-Klein tower of states; if $R$ is sufficiently small, the extra dimensions only become
evident at very high energies.  More recently, it has become popular to consider scenarios
in which Standard Model fields are localized on a brane embedded in a larger bulk
\cite{ArkaniHamed:1998rs,Antoniadis:1998ig,Randall:1999ee,Randall:1999vf}.  In this picture,
the extra dimensions are difficult to perceive because we can't get there.

In this paper we consider a new way to keep extra dimensions hidden, or more generally to
affect the propagation of fields along directions orthogonal to our macroscopic
dimensions:  adding Lorentz-violating tensor fields (``aether'') with expectation values
aligned along the extra directions.  Interactions with the aether modify the dispersion
relations of other fields, leading (with appropriate choice of parameters) to larger energies
associated with extra-dimensional momentum.\footnote{After this paper was completed,
we became aware of closely related work by Rizzo \cite{Rizzo:2005um}. He enumerated a complete 
set of five-dimensional Lorentz-violating operators that preserve Lorentz invariance in 4D, and 
calculated their effect on the spectrum of the Kaluza-Klein tower. In contrast, our starting point
is the expectation value of a dynamical aether field, and its lowest-order couplings to 
ordinary matter.  The modified dispersion relations we derive recover in large measure
Rizzo's phenomenological results.}

This scenario has several novel features.  Most importantly, it allows for completely
different spacings in the Kaluza-Klein towers of each species.  If the couplings are chosen
universally, the extra mass given to fermions will be twice that given to bosons.
There will also be new degrees of freedom associated with fluctuations of the aether field
itself; these are massless Goldstone bosons from the spontaneous breaking of Lorentz
invariance, but can be very weakly coupled to ordinary matter.  There is a sense in which the
effect of the aether field is to distort the background metric, but in a way that is felt differently
by different kinds of fields.  The extra dimensions can 
be ``large'' if the expectation value of the aether field is much larger than the inverse coupling.
In contrast to brane-world models, we expect no deviation from Newton's inverse square law
even if the extra dimensions are as large as a millimeter, 
as the gravitational source will be distributed uniformly throughout the extra dimensions
rather than confined to a brane.  The model has no
obvious connection to the hierarchy problem; indeed, hiding large dimensions requires
the introduction of a new hierarchy.  New physical phenomena associated with the 
scenario deserve more extensive investigation.

\section{Aether}

For definiteness, consider a five-dimensional flat spacetime with coordinates 
$x^a = \{x^\mu, x^5\}$ and metric signature $(-++++)$.  The 
fifth dimension is compactified on a circle of radius $R$.
The aether is a spacelike five-vector $u^a$, and we can define a ``field strength'' tensor
\be
  V_{ab} = \nabla_au_b - \nabla_bu_a\,.
\ee 
This field is not related to the electromagnetic vector potential $A_a$ or its associated
field strength $F_{ab} = \nabla_a A_b - \nabla_b A_a$, nor will the dynamics of $u^a$
respect a U(1) group of gauge transformations.  Rather, the aether field will be
fixed to have a constant norm, with an action
\be
  S = M_*\int d^5x \sqrt{-g}\left[-\frac{1}{4}V_{ab}V^{ab}- \lambda(u_au^a - v^2) + \sum_i\call_i\right]\,,
  \label{ulag}
\ee
The $\call_i$'s are various interaction terms to be considered below, 
and $M_*$ is an overall scaling parameter.  Note that
$\lambda$ is not a fixed parameter, but a Lagrange multiplier enforcing the constraint 
\be
  u^au_a = v^2\,.
  \label{constraint}
\ee
We choose conventions such that $u^a$ has dimensions of mass.
The equation of motion for $u^a$, neglecting interactions with other fields for the moment, is
\be
  \nabla_a V^{ab} + v^{-2}u^b u_c \nabla_d V^{cd} = 0\,.
  \label{eom}
\ee
Any configuration for which $V_{ab} = 0$ everywhere will solve this equation.
In particular, there is a background solution of the form
\be
  u^a = (0,0,0,0,v)\,,
  \label{uexp}
\ee
so that the aether field points exclusively along the extra direction.   We will consider this
solution for most of this paper.

Constraints on four-dimensional Lorentz violation via couplings to Standard Model
fields have been extensively studied \cite{{Carroll:1989vb},{Coleman:1997xq},
{Colladay:1998fq},{Soda:2006wr}}.  The dynamics of the (typically timelike) aether fields
themselves and their gravitational effects
have also been considered \cite{{Kostelecky:1989jw},{Jacobson:2000xp},{Jacobson:2004ts},
{Carroll:2004ai},{Lim:2004js},{Eling:2003rd},{Kostelecky:2003fs},{Bluhm:2007bd}}.  More recently,
attention has turned to the case of spacelike vector fields, especially in the early
universe \cite{Ackerman:2007nb,Dulaney:2008ph}.

The particular form of the Lagrangian (\ref{ulag}) is chosen to
ensure stability of the theory; for spacelike vector fields, a generic set of kinetic terms
would generally give rise to negative-energy excitations.
This specific choice propagates two positive-energy modes: one massless scalar,
and one massless pseudoscalar \cite{Dulaney:2008ph}.  
For purposes of this paper we will not investigate
the fluctuations of $u^a$ in any detail.  Although the modes are light, their couplings to
Standard Model fields can be suppressed.  Nevertheless, we expect that
traditional methods of constraining light scalars (such as limits from stellar cooling) will
provide interesting bounds on the parameter space of these models.

\section{Energy-Momentum and Compactification}

A crucial property of aether fields is the dependence of their energy density on 
the spacetime geometry.  The energy-momentum tensor takes the form
\be
  T_{ab} = V_{ac}V_b{}^c - \frac{1}{4}V_{cd}V^{cd} g_{ab} +
  v^{-2}u_au_bu_c \nabla_d V^{cd}\,.
\ee
In particular, $T_{ab}$ vanishes when $V_{ab}$ vanishes, as for the constant field configuration
in flat space (\ref{uexp}).  The non-vanishing expectation value for the aether field does not
by itself produce any energy density.  In the context of an extra dimension, this implies that
the aether field will not provide a contribution to the effective potential for the radion,
so the task of stabilizing the extra dimension must be left to other mechanisms.

When the background spacetime is not Minkowski, however, even a ``fixed'' aether field
can give a non-vanishing energy-momentum tensor.  In \cite{Carroll:2004ai} it was shown that
a timelike aether field would produce an energy density proportional to the square of the
Hubble constant, while in \cite{Ackerman:2007nb} a spacelike aether field was shown
to produce an anisotropic stress.  We should therefore check that an otherwise
quiescent aether field oriented along an extra dimension
does not create energy density when the four-dimensional
geometry is curved.

Consider a factorizable geometry with an arbitrary four-dimensional metric and a
radion field $b(x^\sigma)$ parameterizing the size of the single extra dimension,
\be
	ds^2 = g_{\mu\nu}(x)dx^\mu dx^\nu + b(x)^2 dx_5^2\,,
\ee
where $x$ here stands for the four-dimensional coordinates $x^\sigma$.
In any such spacetime, there is a background solution
\be
	u^a = \left(0,0,0,0,\frac{v}{b(x)}\right).
	\label{config2}
\ee
It is straightforward to verify that this configuration satisfies the equation of motion
(\ref{eom}), as well as the constraint (\ref{constraint}), even though $V_{ab}$ does not
vanish:
\be
  V_{\mu 5} = - V_{5\mu} = v\nabla_\mu{b}\,.
\ee

We can then calculate the energy-momentum tensor associated with the aether:
\bea
  T^{(u)}_{\mu\nu} &=& \frac{v^2}{b^2}\left(\nabla_\mu b \nabla_\nu b
  -\frac{1}{2} g_{\mu\nu}\nabla_\sigma b \nabla^\sigma b \right)\cr
  T^{(u)}_{\mu 5} &=& 0\cr
  T^{(u)}_{55} &=& v^2 \left(\nabla_\sigma \nabla^\sigma b 
  - \frac{1}{2}\nabla_\sigma b \nabla^\sigma b\right)\,.
\eea
The important feature is that $T^{(u)}_{ab}$ will vanish
when $\nabla_\mu b = 0$.  As long as the extra dimension is stabilized and
the aether takes on the configuration (\ref{config2}), there will be no contributions
to the energy-momentum tensor; in particular, neither the expansion of the universe
nor the spacetime geometry around a localized gravitating source will be affected.

\section{Scalars}

We now return to flat spacetime ($g_{ab} = \eta_{ab}$) and
consider the effects of interactions of the aether on various types of matter
fields, beginning with a real scalar $\phi$.  We impose a $\mathbb Z_2$ symmetry, 
$u^a\rightarrow -u^a$.
The Lagrangian with the lowest-order coupling is then
\be
  \call_\phi = -\frac{1}{2}(\partial\phi)^2  - \frac{1}{2}m^2\phi^2 
  -\frac{1}{2\mu_\phi^2}u^a u^b \partial_a\phi \partial_b\phi\,,
  \label{scalarlag}
\ee
with a corresponding equation of motion
\be
  \partial_a\partial^a\phi - m^2\phi = \mu_\phi^{-2} \partial_a(u^au^b\partial_b\phi)\,. 
  \label{scalareom}
\ee
Expanding the scalar in Fourier modes,
\be
  \phi \propto e^{ik_a x^a} = e^{ik_\mu k^\mu + i k_5x^5}\,,
\ee
yields a dispersion relation
\be
  -k_\mu k^\mu = m^2 + \left(1+ \alpha_\phi^2\right)k_5^2\,,
\ee
where
\be
  \alpha_\phi = {v}/{\mu_\phi}\,.
\ee
Note that with our metric signature, $-k_\mu k^\mu = \omega^2 - \vec{k}^2$.

This simple calculation illustrates the effect of the coupling to the spacelike vector 
field.  Compactifying the fifth dimension on a circle of radius $R$ quantizes the
momentum in that direction, $k_5 = n/R$.  In standard Kaluza-Klein theory, this gives rise
to a tower of states of masses $m^2_{KK} = m^2 + (n/R)^2$.  With the addition of the aether field,
the mass spacing between different states in the KK tower is enhanced,
\be
  m^2_{AC} = m^2 + (1 + \alpha_\phi^2)\left(\frac{n}{R}\right)^2\,.
\ee

The parameter $\alpha_\phi$ is a ratio of the aether vev to the mass scale $\mu_\phi$
characterizing the coupling, and could be much larger than unity.  If the vev is 
$v\sim\mpl$, and the coupling parameter 
is $\mu_\phi\sim$~TeV, the masses of the excited modes are
enhanced by a factor of $10^{15}$.  The extra dimension could be as large as
$R \sim 1$~mm, and the $n=1$ state would have a mass of order TeV.  Admittedly, we
have no compelling reason why there should be such a hierarchy between $v$ and
$\mu_\phi$ at this point, other than that it is interesting to contemplate.

We will examine the effects of aether compactification on gravitons below, but it is
already possible to see that we should not expect any small-scale deviations from Newton's
law, even if the extra dimensions are millimeter-sized.  Unlike braneworld compactifications,
here the sources are not confined to a thin brane embedded in a large bulk; rather, light
fields are zero modes, spread uniformly throughout the extra dimensions.  Therefore, the
gravitational lines of force do not spread out from the source into the higher-dimensional
bulk; the sources are still of \emph{codimension} three in space, and gravity will appear
three-dimensional.  There is correspondingly less motivation for considering 
macroscopic-sized extra dimensions in this scenario, as they would remain undetectable
by tabletop experiments.

One may reasonably ask whether it is appropriate to think of such a scenario as a 
``large'' extra dimension at all, or whether we have simply rescaled the metric in an
unusual way.  In the Lagrangian (\ref{scalarlag}) alone, the effect of the aether field
is simply to modify the metric by a disformal transformation,
$g^{ab} \rightarrow g^{ab} + u^a u^b$.  There is a crucial difference, however, in that
the interaction with the aether vector is generically not universal.  Different fields will
tend to have different mass splittings in their Kaluza-Klein towers.  Indeed, we shall see
that while the splittings for gauge fields follow the pattern of that for scalars, the splittings
for fermions are of order $\alpha^4$ rather than $\alpha^2$, and the splittings for
gravitons do not involve a mass scale $\mu$ at all.  Thus, aether compactification is
conceptually different from an ordinary extra dimension.

Finally, we point out that if we have not imposed the $\mathbb Z_2$ symmetry, the lowest order coupling becomes $\mu^{-1} u^a \partial_a \phi$. By integration by parts, this is equivalent to $-\mu^{-1} (\partial_a u^a)\phi$, which vanishes given our background solution for $u^a$ in (\ref{uexp}).

\section{Gauge Fields}

Consider an Abelian gauge field $A_a$, with field strength tensor $F_{ab}$.
The Lagrangian with the lowest-order coupling to $u^a$ is
\be
  \call_A = -\frac{1}{4}F_{ab}F^{ab} - \frac{1}{2\mu_A^2} u^au^b g^{cd}F_{ac}F_{bd}\,,
  \label{gaugelag}
\ee
with equation of motion
\be
  \partial_aF^{ab} = \mu_A^{-2}\left(u_c u^b \partial_a F^{ca} - u_cu^a\partial_aF^{cb}\right)\,.
  \label{gaugeeom}
\ee
We can decompose this into $b=5$ and $b=\nu$ components in the background 
(\ref{uexp}):
\bea
  \partial_\mu F^{\mu 5}&=&0\,,
  \label{gauge5}\\
  \partial_\mu F^{\mu\nu} &=& -(1+\alpha_A^2)\partial_5 F^{5\nu}\,,
  \label{gaugenu}
\eea
where 
\be
  \alpha_A = {v}/{\mu_A}\,.
\ee

We can take advantage of gauge transformations $A_a \rightarrow A_a + \partial_a \lambda$
to set $A_5 = 0$.
This leaves some residual gauge freedom; we can still transform
$A_\mu \rightarrow A_\mu + \partial_\mu \tilde\lambda$,
as long as $\partial_5\tilde\lambda = 0$.  In other words, the zero mode retains all of
its conventional four-dimensional gauge invariance.

Choose $A_5=0$ gauge, and go to Fourier space,
$A^\nu \propto \epsilon^\nu e^{ik_\mu x^\mu + ik_5 x^5}$,
where $\epsilon^\nu$ is the polarization vector.
Then (\ref{gauge5}) and (\ref{gaugenu}) imply
\bea
  k_5 k_\mu \epsilon^\mu &=& 0\,,
  \label{gaugedisp5} \\
  \left[k_\mu k^\mu + (1+\alpha_A^2) k_5^2 \right]\epsilon^\nu - k^\nu k_\mu \epsilon^\mu&=&0\,.
  \label{gaugedispnu}
\eea
When $k_5=0$, we obtain the ordinary dispersion relation for a photon.
When $k_5$ is not zero, (\ref{gaugedisp5}) implies $k_\mu \epsilon^\mu = 0$,
and the dispersion relation is
\be
  -k_\mu k^\mu = (1+\alpha_A^2) k_5^2\,.
\ee
Precisely as in the scalar case, the Kaluza-Klein masses are enhanced by a factor
$(1+\alpha_A^2)$, although there is no necessary relationship between $\alpha_A$
and $\alpha_\phi$.  The same reasoning would apply to non-Abelian gauge
fields, through a coupling $u^au^b \tr(G_{ac}G_b{}^c)$.

\section{Fermions}

Next we turn to fermions, taken to be Dirac for simplicity.  Given the symmetry 
$u^a \rightarrow -u^a$, we might consider a coupling of the form  
$u^a u^b \bar{\psi}\gamma_a \gamma_b \psi$.
But because $u^au^b$ is symmetric in its two indices, this is equivalent to 
$u^a u^b \bar{\psi}\gamma_{(a} \gamma_{b)} \psi = u^au^b\bar{\psi}g_{ab}\psi
= v^2\bar{\psi}\psi$, so this interaction doesn't violate Lorentz invariance.

The first nontrivial coupling involves one derivative,
\be
  \call_\psi = i \bar{\psi} \gamma^a\partial_a \psi - m\bar\psi \psi 
  - \frac{i}{\mu_\psi^2}u^au^b \bar\psi \gamma_{a}\partial_{b} \psi\,,
  \label{fermilag}
\ee
leading to an equation of motion
\be
  i\gamma^a\partial_a \psi - m \psi 
  - \frac{i}{\mu_\psi^2}u^au^b \gamma_{a}\partial_{b} \psi = 0\,.
\ee
Going to Fourier space as before, we ultimately find a dispersion relation
\be
  -k^a k_a - \frac{2}{\mu_\psi^2} (u^ak_a)^2 - \frac{1}{\mu_\psi^4}
    u^au_a (u^bk_b)^2 = m^2\,.
\ee
Plugging in the background (\ref{uexp}) and defining
\be
  \alpha_\psi = {v}/{\mu_\psi}\,,
\ee
we end up with
\be
  -k^\mu k_\mu = m^2 + (1 + \alpha_\psi^2)^2 k_5^2\,.
\ee

Although the form of this equation is identical to the scalar and gauge-field cases, it
is quantitatively different: for large $\alpha$ the enhancement goes as $\alpha^4$
rather than $\alpha^2$.  If (in the context of some as-yet-unknown underlying theory)
all of the mass scales $\mu$ are similar, we would expect a much larger mass splitting
for fermions in an aether background than for bosons.

Similar to the scalar case, if we do not impose the $\mathbb Z_2$ symmetry, we are led to consider the following two lower order couplings: $u_a \bar{\psi} \gamma^a \psi$ and $\frac{i}{\mu} u^a \bar{\psi} \partial_a \psi$. Following the same procedure as before, the first term leads to the dispersion relation
\be
	-k_{\mu} k^{\mu} = m^2 + v^2 +k_5^2 + 2vk_5 = m^2 + (v+k_5)^2.
\ee 
As usual, coupling to $u^a$ enhances the mass spacing of the KK tower, but now the
spacing will depend on the direction of the 5th-dimensional momentum as well as its magnitude.

Meanwhile, the second term leads to the dispersion
\bea
	-k_{\mu} k^{\mu} &=& m^2 - 2m\alpha k_5 + (1+\alpha^2)k_5^2 \\
	&=& (m-\alpha k_5)^2 + k_5^2,
\eea
where $\alpha = v/\mu$.
Interestingly, if $(1+\alpha^2)/\alpha < 2mR$, this coupling results in a 
reduction in $m^2$ for small $n$. However, it can be checked that these negative mass 
corrections are never sufficiently large to lead to tachyons. For $n$ large, the mass spacing 
is enhanced, as usual. 

\section{Gravity}

The aether field can couple nonminimally to gravity through an action
\be
  S =M_* \int d^5x \sqrt{-g}\left[\frac{\mpl^2}{2}R + \alpha_g u^au^b R_{ab}\right]\,,
  \label{gravitylag}
\ee
where $\mpl$ is the 4-dimensional Planck scale and $\alpha_g$ is dimensionless.
The gravitational equation of motion takes the form
\be
  G_{ab} = \frac{\alpha_g}{2\mpl^2}W_{ab}\,,
  \label{gravityeom}
\ee
where $G_{ab}=R_{ab} -\frac{1}{2}Rg_{ab}$ and
\bea  
  W_{ab} &=& R_{cd}u^cu^dg_{ab} + \nabla_c\nabla_{a}\left(u_{b}u^c\right) +
  \nabla_c\nabla_{b}\left(u_{a}u^c\right) \nonumber \\
  &&\qquad -\nabla_c\nabla_d(u^cu^d)g_{ab} 
  - \nabla_c\nabla^c(u_au_b)\,.
  \label{wab}
\eea

Now we consider small fluctuations of the metric,
\be
  g_{ab} = \eta_{ab} + h_{ab}\,.
\ee
The choice of background field $u^a = (0, 0,0,0, v)$ spontaneously breaks diffeomorphism 
invariance, so not all coordinate transformations are open to us if we want to preserve that form.  
Under an infinitesimal coordinate transformation parameterized by a vector field,
$x^a \rightarrow \bar{x}^a = x^a + \xi^a$, the metric fluctuation and aether change by
$h_{ab} \rightarrow h_{ab} + \partial_a\xi_b + \partial_b\xi_a$ and
$u^a \rightarrow u^a + \partial_5 \xi^a$.
Therefore, we should limit our attention to gauge transformations satisfying
$\partial_5\xi^a = 0$.  We can, for example, set $h_{\mu 5}=0$.
We then still have residual gauge freedom in the form of $\xi^\mu$, as long as
$\partial_5\xi^\mu = 0$.  This amounts to the usual 4-d gauge freedom for the 
massless four-dimensional graviton.

Taking advantage of this gauge freedom, we can partly decompose
the metric perturbation as
\bea
  h_{\mu\nu} &=& \hb_{\mu\nu} + \Psi\eta_\mn\,, \nonumber \\
  h_{55} &=& \Phi\,, 
\eea
where $\eta^{\mn}\hb_\mn=0$.  In this decomposition, $\hb_\mn$ represents 
propagating gravitational waves, $\Psi$ represents Newtonian gravitational fields,
and $\Phi$ is the radion field representing the breathing mode of the extra dimension.
The zero mode of this field is a massless scalar coupled to matter with gravitational
strength; in a phenomenologically viable model, it would have to be stabilized,
presumbably by bulk matter fields.
The Einstein tensor becomes
\bea
  G_\mn &=& \frac{1}{2}\left[-\partial_\lambda\partial^\lambda\hb_\mn -\partial_5^2\hb_\mn
  +\partial_\mu \partial^\lambda \hb_{\lambda\nu} \right. \\
  && \left. + \partial_\nu\partial^\lambda \hb_{\lambda\mu}
  - 2 \partial_\mu \partial_\nu\Psi - \partial_\mu \partial_\nu\Phi  \right. \nonumber \\
  && \left.
  - \left(\partial^\rho \partial^\sigma \hb_{\rho\sigma} 
  - 2\partial_\lambda\partial^\lambda \Psi + 3\partial_5^2\Psi 
  - \partial_\lambda\partial^\lambda\Phi\right)\eta_\mn \right] \,, \label{Gmunu}\nonumber \\
  G_{\mu 5} &=& \frac{1}{2}\left(\partial_5\partial^\lambda \hb_{\lambda\mu}
  -3\partial_\mu\partial_5 \Psi\right)\,, \\
  G_{55} &=& \frac{1}{2} \left(-\partial^\rho\partial^\sigma\hb_{\rho\sigma}
  +3\partial_\lambda\partial^\lambda \Psi\right)\,,
\eea
and (\ref{wab}) is
\bea
  W_\mn &=& v^2\left(\partial_5^2\hb_\mn - 3\partial_5^2\Psi\eta_\mn  
  - \partial_5^2\Phi\eta_\mn\right) \,, \label{Wmunu}\\
  W_{\mu 5} &=& v^2\partial_\mu\partial_5\Phi \,,\\
  W_{55} &=& -2 v^2(2\partial_5^2\Psi + \partial_5^2\Phi)\,.
\eea

We have already argued that there will be no macroscopic deviations from Newton's law
on the scale of the extra-dimensional radius $R$, because the zero-mode fields are
distributed uniformly through the extra dimensions.  However, we can also inquire about
the Kaluza-Klein tower of propagating gravitons.
To that end, we  set $\Phi=0=\Psi$ and consider transverse
waves, $\partial^\lambda \hb_{\lambda\mu}=0$.  The gravity equation (\ref{gravityeom})
becomes
\be
  -\frac{1}{2}\partial_c\partial^c \hb_\mn = \frac{\alpha_g v^2}{2\mpl^2}\partial_5^2\hb_\mn\,.
\ee
This implies a dispersion relation
\be
  -k_\mu k^\mu = \left(1 + \frac{\alpha_g v^2}{\mpl^2}\right)k_5^2\,.
\ee
As before, there is a altered dispersion relation for modes with bulk momentum.
However, the dimensionless coupling $\alpha_g$ appears directly in the Lagrangian,
rather than arising as a ratio $\alpha = v/\mu$.  It is therefore consistent to imagine
scenarios with $\alpha_g\sim 1$, while the other $\alpha_i$'s are substantially larger.
In that case, KK gravitons will have masses that are close to the conventional
expectation, $m\approx n/R$, even while other fields are much heavier.
In the scenario with a single extra dimension, the underlying quantum-gravity scale
$M^3_{QG} = M_*M_P^2$ will still be substantially larger than a TeV, and we do not
expect graviton production at colliders; but such a phenomenon might be important
in extensions with more than one extra dimension.

\section{Conclusions}

The presence of Lorentz-violating aether fields in extra dimensions introduces novel
effects into Kaluza-Klein compactification schemes.  Interactions with the
aether alter the relationship between the size of the extra dimensions
and the mass splittings within the KK towers.  With appropriately chosen
parameters, modes with extra-dimensional momentum can appear very heavy from a
four-dimensional perspective, even with relatively large extra dimensions.

A number of empirical tests of this idea suggest themselves.  The most obvious is the
possibility of KK towers with substantially different masses for different
species.  While scalar and gauge-boson mass splittings follow a similar pattern, 
fermions experience greater enhancement, while gravitons can naturally be
less massive.  In addition, although we have not considered the prospect carefully in this
paper, oscillations of the aether field itself are potentially detectable.  Their couplings
will be suppressed by the mass scales $\mu_i$, without being enhanced by the vev
$v$; nevertheless, searches for 
massless Goldstone bosons should provide interesting constraints on the parameter
space.

Our investigation has been phenomenological in nature; we do not have an
underlying theory of the aether field, nor any natural expectation for the magnitudes of
the parameters $v$, $\mu_i$, and $\alpha_g$.  The possibility of a hidden
millimeter-sized dimension requires a substantial hierarchy, $v/\mu_i \sim 10^{15}$;
even in the absence of such large numbers, however, interactions with the aether may
lead to subtle yet important effects.
It would certainly be interesting to have a deeper understanding of the possible 
origin of these fields and couplings.

Numerous questions remain to be addressed.  We considered a vector field
in a single extra dimension, but higher-rank tensors in multiple dimensions should lead
to analogous effects.  It would also be interesting to study the gravitational effects
of the aether fields themselves in non-trivial spacetime backgrounds.  The idea of
modified extra-dimensional dispersion relations in the presence of Lorentz-violating
tensor fields opens up a variety of possibilities that merit further exploration.

%\section*{Acknowledgments}

We are very grateful to Mark Wise for helpful conversations.  This research was supported in 
part by the U.S. Department of Energy and by the David and Lucile Packard Foundation.

\end{document}